\documentclass[journal,lettersize]{IEEEtran}

\usepackage{amsmath,amsfonts}
\usepackage{algorithmic}
\usepackage{algorithm}
\usepackage{array}
\usepackage[caption=false,font=normalsize,labelfont=sf,textfont=sf]{subfig}
\usepackage{textcomp}
\usepackage{stfloats}
\usepackage{url}
\usepackage{verbatim}
\usepackage{graphicx}
\usepackage{cite}
\usepackage{xcolor}

\begin{document}

\title{Towards Precise Channel Knowledge Map: Exploiting Environmental Information from 2D Visuals to 3D Point Clouds}

\author{Yancheng Wang,~\IEEEmembership{Member,~IEEE}, Chuan~Huang,~\IEEEmembership{Member,~IEEE}, 
Songyang~Zhang,~\IEEEmembership{Member,~IEEE},
Guanying~Chen,~\IEEEmembership{Member,~IEEE}, 
Wei~Guo,~\IEEEmembership{Member,~IEEE},
Shenglun~Lan,
Lexi~Xu,~\IEEEmembership{Senior~Member,~IEEE}, Xinzhou~Cheng,
Xiongyan~Tang,~\IEEEmembership{Senior~Member,~IEEE}, and Shuguang~Cui,~\IEEEmembership{Fellow,~IEEE} 
\thanks{The work was supported in part by the key technology project of Shenzhen with Grant No. KJZD20230923115104009. \emph{(Corresponding authors: Chuan Huang and Wei Guo.)} 

Y. Wang is with the Shenzhen Institute for Advanced Study, University of Electronic Science and Technology of China (UESTC), the School of Science and Engineering, The Chinese University of Hong Kong, Shenzhen, China; C. Huang is with the Shenzhen Institute for Advanced Study, University of Electronic Science and Technology of China (UESTC), and the Shenzhen Future Network of Intelligence Institute, Shenzhen, China. S. Cui is with the School of Science and Engineering, the Shenzhen Future Network of Intelligence Institute, and the Guangdong Provincial Key Laboratory of Future Networks of Intelligence, The Chinese University of Hong Kong, Shenzhen, China; S. Zhang is with University of Louisiana at Lafayette, USA; G. Chen is with Sun Yat-sen University, Shenzhen, China; W. Guo is with The Hong Kong University of Science and Technology, Hong Kong SAR, China; S. Lan is with Changchun University of Science and Technology, China; L. Xu, X. Cheng, and X. Tang are with the Research Institute, China Unicom, Beijing, China.
}
}

\maketitle

\begin{abstract}
The substantial communication resources consumed by conventional pilot-based channel sounding impose an unsustainable overhead, presenting a critical scalability challenge for future 6G networks characterized by massive channel dimensions, ultra-wide bandwidth, and dense user deployments. As a generalization of radio maps, the channel knowledge map (CKM) offers a paradigm shift, enabling access to location-tagged channel information without exhaustive measurements. To fully utilize the potential of CKMs, this work highlights the necessity of leveraging three-dimensional (3D) environmental information, beyond conventional two-dimensional (2D) visual representations, to construct high-precision CKMs. Specifically, we present a novel framework that integrates 3D point clouds into CKM construction through a hybrid model- and data-driven approach, with extensive case studies in real-world scenarios. The experimental results demonstrate the potential for constructing precise CKMs based on 3D environments enhanced with semantic understanding, as well as their applications in next-generation wireless communications. We also release a real-world dataset of measured channels paired with high-resolution 3D environmental data to support future research and validation.

\end{abstract}


\section{Introduction}

The forthcoming sixth-generation (6G) wireless systems are expected to enable a wide range of breakthrough applications, ranging from extended reality (XR) and space-air-ground integrated communications to connected and autonomous vehicles (CAV), which impose stringent performance requirements, such as peak terabit-per-second data rates, sub-millisecond latency, high-precision synchronization, and extreme reliability~\cite{tataria20216g}. To achieve these goals, an extensive deployment of ultra-massive multiple-input multiple-output (MIMO) antenna systems is demanded, potentially involving hundreds or even thousands of antennas, along with significantly wider bandwidths and operation at higher frequency bands.
However, conventional methods of acquiring channel state information (CSI), predominantly through pilot-based measurements, are becoming increasingly unsustainable for such high-dimensional channels due to their excessive resource consumption and limited scalability, especially in wideband and ultra-massive MIMO systems~\cite{zeng2021toward,zeng2023tutorial}. 

\begin{figure*}
\centering
\includegraphics[width=0.8\linewidth]{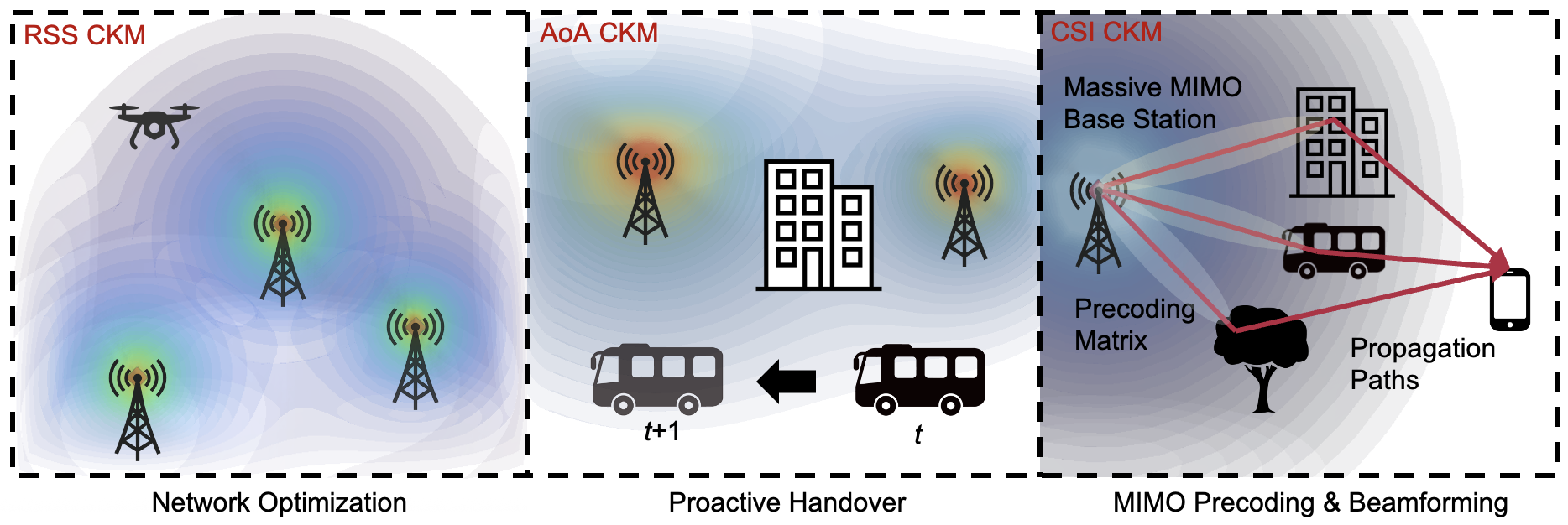}
\caption{{Representative applications of CKM}: 1) \textbf{Network Optimization and Planning}. Received signal strength (RSS) CKMs enable spectrum management, coverage optimization, and interference control; in low-altitude economies, drones rely on 3D CKMs to avoid coverage holes. 2) \textbf{Proactive Handover}. Angle of arrival (AoA) CKMs forecast imminent blockages (e.g., by buildings) and enable seamless mobility-aware handovers, such as in vehicular communications. 3) \textbf{MIMO Precoding and Beamforming}: Channel state information (CSI) CKMs support position-assisted beam selection and training, reducing pilot-signal overhead and ensuring robust link performance.}\label{f_usecase}
\end{figure*}

To address these limitations, channel knowledge map (CKM) has emerged as an efficient tool for representing channel information, offering a promising solution to characterizing complex spectrum environments~\cite{zeng2021toward}. In general, CKM stores the channel parameters in an area of interest (AoI), providing fine-grained, location-specific information on wireless channels. Typical characteristics include received signal strength (RSS), angle of arrival (AoA)/departure (AoD), channel impulse response (CIR), power delay profile (PDP), and CSI~\cite{zeng2023tutorial,wang2024channel,wang2025point}. 
Unlike traditional channel estimation that relies on instantaneous measurements, CKMs aim to generalize the representation across space and time, reducing pilot overhead and thus showing significant transformative potential in next-generation wireless communications. Typical applications include CKM-assisted network optimization, proactive service and application adaptation, and MIMO beamforming. For example, in ultra-massive MIMO systems, CKMs facilitate training-free adaptation by preselecting dominant propagation paths or initializing precoding vectors without exhaustive beam sweeping~\cite{zeng2023tutorial}, improving link establishment latency and reducing complexity, as shown in Fig.~\ref{f_usecase}. To leverage the power of CKMs, the first step is to efficiently obtain a precise CKM, making CKM construction a critical problem in modern wireless networking.

To date, various methods have been proposed to construct CKMs, each with inherent limitations.
\emph{Classical interpolation methods} such as Kriging and tensor completion~\cite{chowdappa2018distributed} rely on spatial correlations.
However, these approaches require abundant channel samples, which are often hard to obtain and fail to capture abrupt changes in signal within obstructed environments.
\emph{Ray-tracing methods}, including recent point cloud-based ray-tracing, simulate the CKM based on detailed environmental models and difficult-to-obtain material parameters~\cite{da2021ray}.
Although they generally achieve high accuracy, these methods suffer from an expensive computational burden and inevitable modeling inaccuracies, making them unsuitable for real-time network management.
Recent interest in CKM construction is attracted to \emph{AI-based approaches}, which learn mappings from environmental characteristics to channel measurements.
Typical examples include convolutional neural networks (CNNs), generative adversarial networks (GANs)~\cite{zhang2023rme}, and RadioUNet~\cite{levie2021radiounet}, which operate on 2D visuals.
More recently, neural rendering techniques have been explored for wireless modeling via implicit or explicit 3D scene representations, such as neural radiance fields (NeRF) and 3D Gaussian splatting (3DGS)~\cite{wen2025wrfgs}, thereby tightly coupling 3D geometry with data-driven learning.

To achieve accurate and robust CKM reconstruction, it is essential to go beyond traditional 2D visuals and utilize 3D environmental information~\cite{wang2024channel,wang2025point}.
Since the wireless channel is inherently determined by the surrounding 3D environment, using 2D maps, which are projections of the 3D world, fails to capture critical features such as elevation, curvature, and obstructing objects.
For example, a low-rise building may seem to block a high-rise base station (BS) in 2D, while 3D reveals a clear line-of-sight.
Additionally, many CKM applications inherently require 3D information.
Optimizing coverage in low-altitude wireless networks, beamforming with elevation angle domain, and MIMO precoding all depend on CKM in the 3D space, making 3D environmental information critical to the construction of 3D CKMs.

This work presents a systematic environment-aware CKM reconstruction framework~\cite{wang2024channel,wang2025point} based on detailed 3D point clouds, capturing fine-grained structural and semantic environmental features; thereby enabling more precise and robust CKM reconstruction. Specifically, we first revisit the CKM construction problem and classify CKMs into three categories. We then briefly review recent advances in 3D environment reconstruction and semantic scene understanding. Building on these advances, we present a hybrid model- and data-driven framework that leverages 3D point clouds for CKM reconstruction. Finally, we outline key challenges and opportunities, and introduce a real-world dataset comprising measured CKMs and reconstructed 3D environment.

\section{CKM Reconstruction: Formulation and Challenges}

The CKM is conceptualized as a geo-tagged, site-specific database designed to represent the intrinsic radio propagation characteristics of a local environment~\cite{zeng2023tutorial}. %Emerging as a critical enabler for environment-aware 6G networks, the CKM paradigm addresses the prohibitive overhead and latency associated with real-time CSI estimation, particularly in future systems with large-dimensional channels and dense node deployment. 
In this section, we first overview the preliminaries of CKM construction.

\subsection{Category of CKMs}
Unlike conventional categorization of CKMs, we introduce a more intuitive classification based on the dimensionality of the stored channel knowledge: namely, {\textit{scalar-valued CKMs}, \textit{vector-valued CKMs}, and \textit{tensor-valued CKMs}}. To maintain clarity, we focus primarily on the canonical BS-to-any (B2X) setting with a fixed transmitter, where the CKM construction is performed on the infrastructure side to minimize user-side overhead.

\begin{itemize}
    \item \textbf{Scalar-valued CKMs:} Scalar-valued CKMs store large-scale fading parameters at each location on the map. Some important cases include path loss, RSS, and signal-to-interference-plus-noise ratio (SINR). The scalar-valued CKMs are in a similar format of gray-scale images, where each pixel is annotated with the corresponding channel parameters, indicating link quality and interference. This kind of CKM provides important information for applications, such as network optimization, spectrum management, and interference coordination.

    \item \textbf{Vector-valued CKMs:} Each location in a vector-valued CKM is annotated with a one-dimensional array of channel parameters, such as PDP, CIR, angular power spectrum (AoA/AoD), and power spectrum. These vector-valued knowledge reveal the underlying geometric characteristics of multipath propagation, including dominant directions and path delays, as well as the complete CIR of a single-input single-output (SISO) channel. Vector-valued CKMs provide a richer characterization of the channel than scalar-valued maps and are well suited for applications such as beam training, proactive handover, and user mobility prediction.
    \item \textbf{Tensor-valued CKMs:} Tensor-valued CKMs store high-dimensional channel knowledge at each location. For example, in a MIMO system, each point in the space can be associated with a full complex-valued CSI matrix $\mathbf{H} \in \mathbb{C}^{N \times M}$ for an N-to-M antenna system, or a 3D tensor $\mathbf{H} \in \mathbb{C}^{N \times M \times F}$ for a MIMO-OFDM system with $F$ denoting the number of subcarriers. Tensor CKMs provide a near-complete representation of the wireless channel and are theoretically sufficient for eliminating the pilot.
\end{itemize}

This categorization of CKMs based on data dimension facilitates the selection of appropriate representations for CKM data format, depending on the requirements of application layers and system complexity.

\subsection{CKM Construction}

In general, constructing a CKM database is formally defined as learning a mapping function $\mathbf{H}_i = f(\mathbf{x}_i | \mathcal{I})$ to complete the missing knowledge across the AoI for any receiver location $\mathbf{x}_i \in \mathbb{R}^3$. The inference is conditioned on the set of available information $\mathcal{I}$, which comprises any of three key inputs: (i) sparse ground-truth measurements $\mathcal{D}$; (ii) a 2D or 3D representation of the physical environment $\mathcal{E}$; and (iii) transmitter location $\mathbf{x}_{\mathrm{T}}$, optionally with its antenna configuration $\mathcal{A}$. The goal is to build an estimator $\hat{f}$ that leverages these information to accurately estimate $\mathbf{H}_i$ for all locations $\mathbf{x}_i$ within the AoI~\cite{zeng2023tutorial}. 

The methodology for constructing such CKM must balance the trade-off between computationally expensive high-fidelity simulators (e.g., ray tracing) and more efficient but data-sensitive AI-based models. 
In this article, we take the construction of scalar-valued and vector-valued CKMs as examples, which will be discussed in Section \ref{sec: case}. Specifically, we target the prediction of the RSS (scalar-valued) and PDP (vector-valued), which is the magnitude of the discrete-time CIR. We note that the fidelity required for complex tensor-valued CKMs, particularly for millimeter-wave scenarios, may still necessitate physics-based simulations.

\subsection{Challenges of CKM Construction with 3D Environments}
Unlike CKM construction from 2D visuals, accurate 3D-based CKM construction remains an ill-studied problem due to the trade-off between complexity and resolution. Ray-tracing methods estimate the channel using the physical environment $\mathcal{E}$, transmitter location $\mathbf{x}_{\mathrm{T}}$, and antenna configuration $\mathcal{A}$, but they are computationally intensive and sensitive to errors in geometry or material properties. Moreover, relying on physical models, these methods struggle to incorporate real-world measurements $\mathcal{D}$. On the other hand, existing data-driven approaches rely on 2D projections $\mathcal{E}_{2D}$ (e.g., images or floor plans) together with measurements $\mathcal{D}$ to predict the channel. Despite some successes, these methods simplify the geometry and hinder accurate modeling of radio propagation behavior in 3D environments, reducing the resolution and precision of CKM construction.

To address these challenges, we propose a hybrid framework~\cite{wang2024channel,wang2025point}, integrating physical models and data-driven neural networks, that directly utilizes high-fidelity 3D environmental representations in the form of point clouds.
Our key innovation is a hybrid paradigm that uses 3D geometry as a physical prior to guide the network, thereby bypassing heavy intersection computations while maintaining generalizability through measured-channel supervision.

\section{3D Environment Reconstruction}

Wireless signal propagation is strongly influenced by the geometry and material properties of the surrounding environment, which govern signal reflection, scattering, and attenuation. Therefore, constructing a high-fidelity 3D model annotated with material semantics is essential for accurate 3D radio map generation.

\subsection{3D Geometry Reconstruction}

Recent advances in 3D reconstruction have enabled high-fidelity modeling of complex environments. Common representations include point clouds, meshes, voxels, NeRF, and 3DGS~\cite{kerbl20233d}. Among these, point clouds offer an effective trade-off between geometric precision and computational efficiency, making them well-suited for large-scale urban radio map construction. 
Additionally, point clouds can be derived from other formats, allowing flexible integration across modalities.

To build accurate 3D maps of urban environments, we adopt a multi-source data fusion strategy that combines drone-captured multi-view images with aerial LiDAR scans. This hybrid approach leverages the complementary strengths of both modalities: images provide rich texture and appearance cues, while LiDAR supplies precise depth measurements and robust coverage in low-texture or occluded areas.

The reconstruction pipeline proceeds as follows. First, Structure-from-Motion (SfM) is applied to the drone images to estimate camera poses and a sparse 3D point cloud. Next, Multi-View Stereo (MVS) techniques are used to densify the point cloud by enforcing photometric consistency across views—minimizing color discrepancies between projections of the same 3D point. Simultaneously, LiDAR point clouds are aligned with the image-based reconstruction through rigid registration and used as geometric priors. These priors help mitigate ambiguities in textureless regions and correct for drift in image-only reconstructions.

\subsection{Semantic Understanding of Environment}

Beyond geometry, signal propagation is heavily influenced by the material properties of surrounding surfaces. For instance, buildings with metal or concrete surfaces strongly reflect or block radio signals; vegetation introduces significant attenuation and scattering; and roads or open grounds often facilitate line-of-sight transmission. Accurately modeling these material interactions is crucial for reliable wireless coverage prediction and radio map construction.

To capture this, we augment the reconstructed 3D environment with semantic and material labels. Each 3D point or surface patch is classified into categories such as \textit{building}, \textit{road}, \textit{tree}, \textit{vegetation}, and \textit{vehicle}, providing structured insight into how different regions interact with radio signals.

We employ modern computer vision models to automatically segment image regions into object and material categories from RGB images~\cite{kirillov2023segment}. Recent advances in open-set semantic segmentation further allow flexible, general-purpose labeling of novel object types without the need for retraining, which is particularly useful for adapting to diverse urban environments.

To achieve accurate 3D semantic understanding, we fuse 2D per-image semantic labels into the 3D reconstruction by training a semantic-aware NeRF~\cite{zhang2024aerial}. Each image is first semantically segmented, and the resulting per-pixel labels are projected into 3D space along camera rays. We extend the NeRF model to jointly predict both color and semantic class probabilities at each 3D location. During training, rendered semantic predictions are supervised by the 2D labels, enforcing cross-view consistency. This yields a 3D representation that is both geometrically accurate and semantically consistent across views.
Incorporating these semantic priors into the 3D model significantly enhances the accuracy and interpretability of electromagnetic simulations and wireless channel modeling, enabling more effective network planning and signal optimization in complex environments.

\section{Point Cloud-Based CKM Construction Method}

\begin{figure*}[!ht]
\centering
\includegraphics[width=0.9\linewidth]{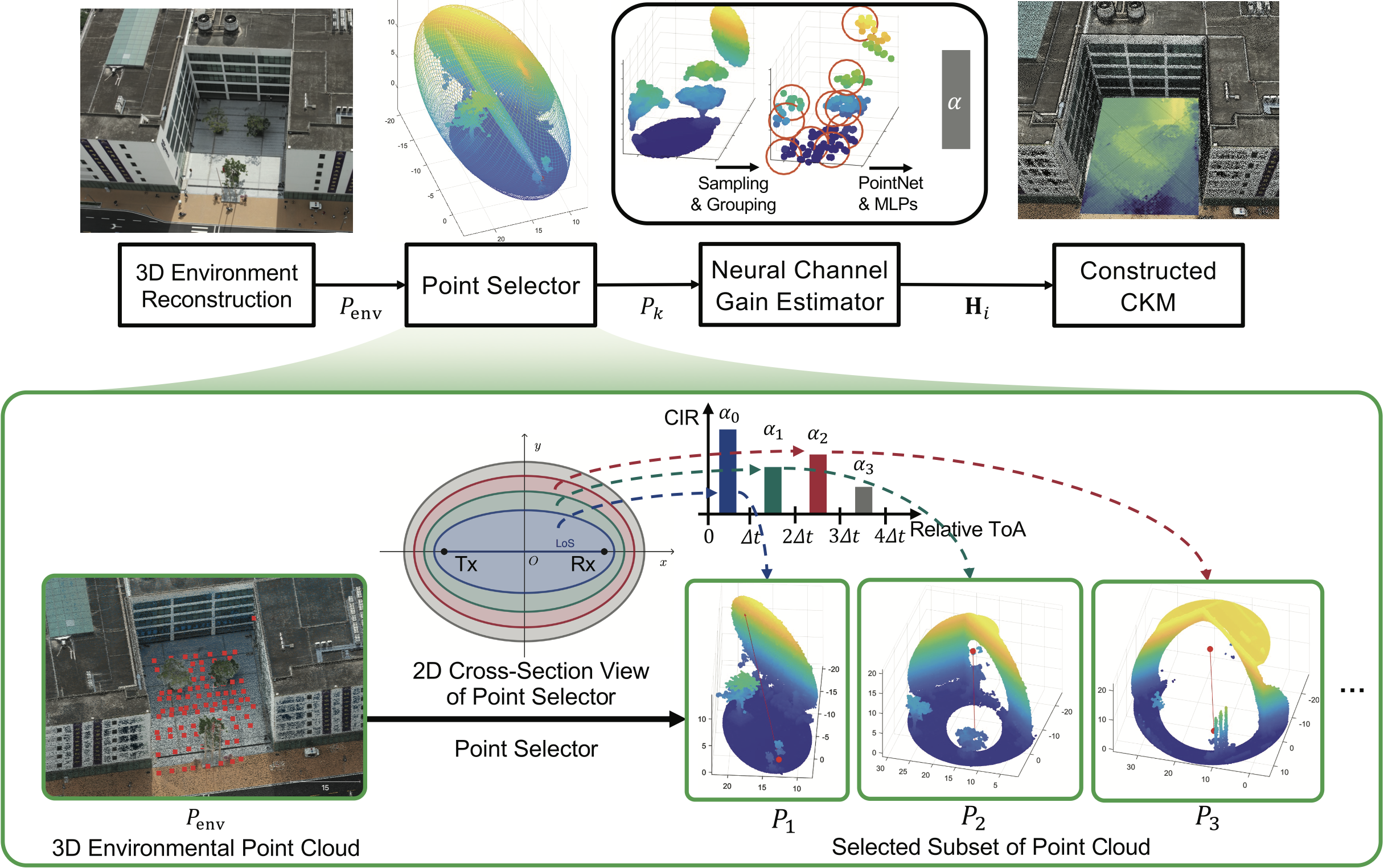} 
\caption{Flowchart of the proposed CKM construction method. The Point Selector partitions the environment into mutually exclusive regions. Each region is defined by the space between two confocal ellipsoids with the Tx and Rx as foci. Points within a specific region correspond to EM waves arriving within the same ToA bin.}
\label{f_flow}
\end{figure*}

With the reconstructed 3D environment, we now describe the proposed method for CKM construction~\cite{wang2025point,wang2024channel}, which efficiently combines the high-resolution point clouds with empirical wireless measurements to enhance prediction accuracy. 

\subsection{Two Key Modules}

Shown as Fig.~\ref{f_flow}, the framework contains two key modules: 1) Point Selector to select propagation-related points from the entire 3D environmental point cloud, and 2) Neural Channel Gain Estimator to predict channel gain using the selected points, detailed as follows.

\subsubsection{{Point Selector}}

The key challenge in leveraging point clouds for wireless channel modeling is scale: a realistic 3D environmental point cloud $\mathcal{P}_\text{env}$ can contain millions of points, yet only a small fraction meaningfully affects the received signal at a given time of arrival (ToA). Processing the entire point cloud not only wastes computation, but also dilutes the model with irrelevant features. Moreover, standard point cloud compression methods such as voxelization aggregate local structures within grid cells, which may blur fine-grained geometric details. To this end, the Point Selector module introduces a model-driven method to filter the raw point cloud, ensuring that only the geometrically relevant points are preserved for subsequent learning\cite{wang2025point,wang2024channel}. 

The design is inspired by the geometry of ellipsoids. By setting the transmitter and receiver as the foci of an ellipsoid, an ellipsoid can be drawn such that every point on its surface corresponds to a constant propagation distance. This property naturally links spatial regions of the environment to a specific ToA. To account for the finite resolution of the discrete channel model, two confocal ellipsoids are defined to represent the lower and upper bounds of a ToA bin. The space between them contains all points that may scatter, reflect, or block EM waves arriving at the receiver within that ToA interval. Filtering the raw environment point cloud through this confocal ellipsoid shell yields a selected subset of point cloud, $\mathcal{P}_k$, which is physically tied to the channel response at time $t_k$~\cite{wang2025point}.

This design has three advantages. First, it establishes a direct, interpretable link between environment geometry and channel response. Second, it dramatically reduces the input size for subsequent processing, shrinking the data from millions of points to a small, meaningful subset while still retaining high-resolution geometry and semantic labels. Finally, it bypasses the heavy computations of ray tracing, enabling a scalable solution for large environments.

\subsubsection{Neural Channel Gain Estimator}

The channel gain estimator is designed as a neural network, where a hierarchical feature extraction strategy is adopted, inspired by PointNet++~\cite{qi2017pointnetplusplus}.
For input, the point cloud is formatted into a matrix in which each row represents a point described by its spatial coordinates augmented with auxiliary features such as color, semantic labels, or geometric attributes, providing valuable context for EM propagation~\cite{wang2025point}. The output is the corresponding CKM values at target locations.

The estimator captures fine-grained local geometric patterns through sampling, grouping, and multi-layer perceptron (MLP) layers, then progressively aggregates them into a global feature vector that encodes both the scene and channel propagation-related properties~\cite{wang2025point}. This vector is passed through an MLP to estimate the final channel gain. Other optional backbones for this module include transformer-based networks, which use self-attention mechanisms to dynamically learn relationships among points in the propagation.

Integrating physics-informed models with data-driven neural networks offers a powerful way to exploit environmental information more efficiently and deeply. Such integration provides neural networks with highly informative inputs, enabling them to go beyond purely statistical learning and capture essential physical characteristics of wireless propagation. Looking ahead, building CKMs that work reliably in real and complex environments will require combining the insights of models with the generalizability of machine learning.

\subsection{Case Studies} \label{sec: case}

\begin{figure}
\centering
\includegraphics[width=1\linewidth]{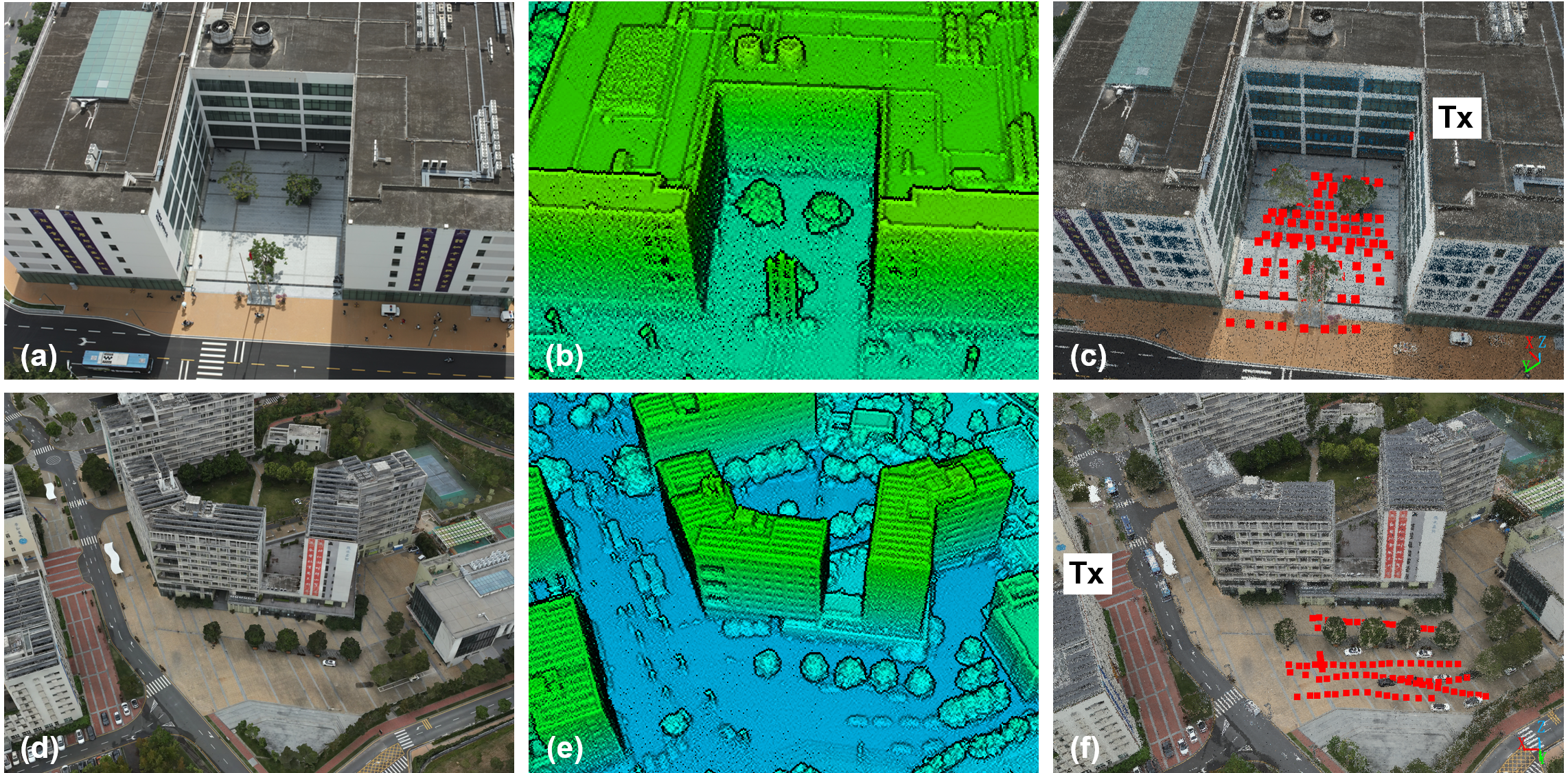}
\caption{Illustration of channel sounding sites and the point clouds. First row: the first AoI, located at the lower campus of CUHK-Shenzhen. Second row: the second AoI, located at the upper campus of CUHK-Shenzhen. For each AoI, from left to right: 1) RGB images of the environment; 2) point cloud; and 3) locations of the channel measurements.}\label{f_dataset}
\end{figure}

To showcase the efficacy and robustness of our proposed CKM construction method, two representative cases are investigated: vector-valued CKM reconstruction and scalar-valued CKM reconstruction. Experiments were conducted on the outdoor channel dataset collected on the campus of The Chinese University of Hong Kong, Shenzhen, as shown in Fig.~\ref{f_dataset}. 

\subsubsection{Vector-Valued CKM Reconstruction}

\begin{figure*}[!t]
    \centering
    \includegraphics[width=1\linewidth]{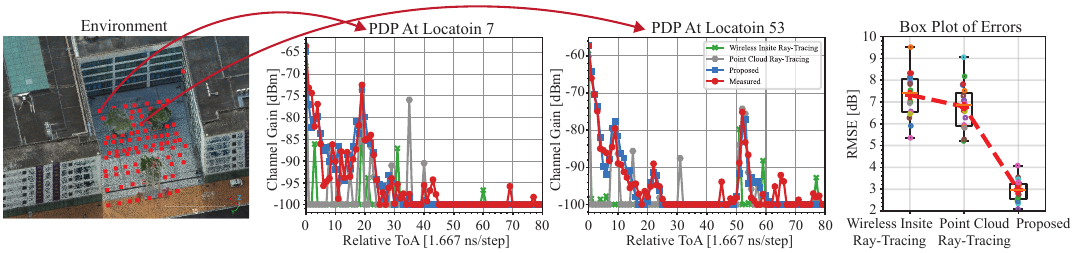}
    \caption{Results of vector-valued CKM reconstruction: The compared ray-tracing methods can accurately identify the ToA of most propagation paths but introduce additional channel gain errors and false paths; The proposed method predicts the PDP more accurately through a data-driven approach.}
    \label{F_cir_measure}
\end{figure*}

Vector-valued CKMs provide rich multipath channel information. In this case study, we evaluate the proposed method in predicting the PDP vector. We compare three approaches against ground-truth measurements: (i) the Wireless InSite ray-tracing software (\url{www.remcom.com/wireless-insite-propagation-software}), (ii) a point cloud-based ray-tracing method, and (iii) the proposed method. The measured dataset consists of CIR sequences collected at 96 locations in a 3D urban canyon-like environment, as shown in Fig.~\ref{F_cir_measure}. We use 80\% of the receiver locations for training and the remaining 20\% for testing. The scene contains dense scatterers, trees, and three-sided concrete-glass buildings, consistent with the 3GPP urban canyon model. 

Among the evaluated approaches, the proposed method achieves the lowest mean prediction error of 2.95~dB, significantly outperforming commercial ray-tracing (7.32~dB) and point cloud ray-tracing (6.79~dB). Fig.~\ref{F_cir_measure} shows that the predicted PDPs better match the measurements across test locations. 

While ray-tracing methods accurately capture the ToA for major paths, they often exhibit deviations in channel gain. As a deterministic method, ray-tracing is highly sensitive to material parameters and geometric modeling, which may fail to account for glass modeling defects, foliage, or surface irregularities. These modeling gaps often lead to spurious paths or inaccurate absorption and non-specular reflection losses. In contrast, our hybrid approach utilizes measured-channel supervision to learn and correct these physical modeling errors. By training on site-specific data, the neural network effectively compensates for the lack of precise physical descriptors for irregular surfaces, achieving higher accuracy at the cost of being data-dependent, whereas ray-tracing remains a zero-shot but less precise alternative.

\subsubsection{Scalar-Valued CKM Reconstruction}

\begin{figure}
    \centering
    \includegraphics[width=1\linewidth]{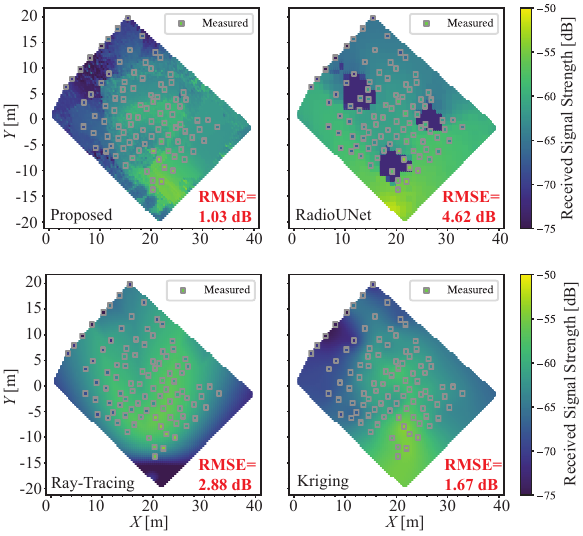}
    \caption{Scalar-valued CKM reconstruction results: The proposed method demonstrates the highest consistency with measured data; CNN-based RadioUNet tends to overestimate path loss in vegetated areas due to reliance on 2D projections; Ray-tracing performance degrades in tree-blockage regions; Kriging performs relatively well but its accuracy is limited without environmental information.}
    \label{F_RM_result_zx}
\end{figure}

We also evaluate the performance to reconstruct the traditional radio map of RSS.
As shown in Fig.~\ref{F_RM_result_zx}, the proposed method achieves the lowest RMSE among all four approaches. Specifically, conventional CNN-based methods, such as RadioUNet, tend to overestimate path loss under vegetated areas due to their 2D reliance. Kriging interpolation performs competitively by leveraging measurement data, while ray-tracing shows moderate errors due to oversimplified trees and imprecise material.

These results emphasize the necessity of using a hybrid approach and incorporating 3D environments. Model-driven ray-tracing struggles in complex outdoor scenarios with inaccurate material parameters, data-driven interpolation offers only partial improvements without exploiting environmental context, and 2D-based AI methods suffer performance degradation in 3D environments. By combining a model-driven module for processing 3D point clouds with a data-driven learning module, the proposed approach achieves the most accurate and robust scalar-valued CKM reconstruction.

\section{Challenges and Opportunities}

To facilitate practical deployment of CKMs in next-generation wireless networks, several deployment-oriented considerations, challenges, and future directions are discussed.

\subsection{Effective Integration of Physics and AI Models}  

Integrating physics-based models with AI provides a powerful way to balance prediction accuracy, interpretability, and runtime efficiency in CKM deployment. Beyond the discussed framework, this integration can be advanced by embedding physical mechanisms directly into learning models, for example through physics-inspired feature extraction, propagation-model-based loss functions, or network architectures constrained by physical laws~\cite{zhang2024physics}. In addition, neural scene representation methods such as NeRF and Gaussian splatting offer promising potential for more expressive and efficient CKM inference.

\subsection{Cross-site Deployment and Zero-Shot Construction} 
In practical applications, site-specific factors, such as material properties and building layouts, often cause sharp variations in channel characteristics. Thus, a key frontier in CKM research is achieving robust generalization across diverse environments without extensive retraining. This challenge can be addressed by transfer/meta-learning methods that extract environment-invariant knowledge and adapt models to new scenarios with limited data. In addition, constructing an EM knowledge base that stores the propagation characteristics of typical objects can further enable cross-site deployment with few-shot calibration in previously unseen environments.

\subsection{Map Refresh and Runtime Update} 

Wireless environments are inherently time-varying, leading to a demand for timely CKM refresh to capture dynamics such as moving users, vehicles, and scattering objects.
Addressing this issue requires efficient scene modeling and low-latency update mechanisms.
Potential solutions include leveraging sensing modalities, such as LiDAR, to detect dynamic objects and rapidly feed their positions into the CKM pipeline.
Another promising direction is to develop efficient CKM inference architectures that can adapt to scene dynamics.
Exemplary approaches include point cloud-based models or Gaussian splatting representations.
Moreover, model compression techniques, such as pruning and split learning, can further reduce the runtime and computational overheads in practical deployment.

\subsection{Positioning Requirements and Localization} 
CKMs are geo-tagged and require accurate positioning and coordinate registration; errors propagate to CKM queries and degrade beam selection and throughput~\cite{zeng2021toward}. 
In practical deployments, GNSS/RTK and UWB provide sub-meter-level accuracy, which is sufficient for many scenarios, while further research into higher-precision positioning remains beneficial.
Conversely, CKMs also enable inverse localization and environment inference through fingerprint matching and advanced RF sensing that exploits multipath and angular features to estimate reflector and scatterer locations for environment reconstruction.

\subsection{Available Data for Evaluation}
Collecting real-world CKM data is expensive and complex, as it requires high-precision instruments (e.g., GHz-bandwidth vector signal generators and spectrum analyzers) and accurate localization devices (e.g., RTK for centimeter-level outdoor positioning). Existing public datasets are mostly synthetic, limited to RSS values, and lack joint channel-environment measurements. To address this gap, we provide a campus outdoor dataset (\url{https://github.com/ycw671/CKM-Dataset}) with extensive measurement points, high temporal resolution (0.4 ns), and jointly measured CIRs, PDPs, RSS values, and 3D environmental point clouds. Looking ahead, the establishment of more diverse CKM datasets covering different 3D environments emerges as a critical task, particularly for supporting applications in the low-altitude economy such as drone communications.

\section{Conclusion}
This article provides an overview of high-fidelity CKM construction utilizing high-dimensional environmental information, highlighting the transformation from simple 2D visuals to high-precision 3D point clouds. Specifically, we first overview the efficient construction of 3D point clouds in wireless systems, after which a hybrid model- and data-driven framework is introduced for CKM construction.
By leveraging detailed 3D geometry and material semantics, our approach enables more realistic modeling of signal propagation, bridging the gap between synthetic assumptions and real-world complexity. As wireless networks become increasingly adaptive and environment-aware, point cloud-based CKM construction offers a promising pathway toward more robust, efficient, and intelligent connectivity, which is expected to stimulate more exciting wireless techniques and significantly benefit future wireless communication networks.

\bibliographystyle{IEEEtran}
\bibliography{reference,IEEEabrv}

\section*{Biography}

\noindent\textbf{Yancheng Wang} is a Ph.D. candidate in the School of Science and Engineering at The Chinese University of Hong Kong, Shenzhen, China.

\vspace{1mm}
\noindent\textbf{Chuan Huang} received the Ph.D. degree from Texas A\&M University, College Station, TX, USA. He is currently a Professor with the Shenzhen Institute for Advanced Study, University of Electronic Science and Technology of China, and the Shenzhen Future Network of Intelligence Institute.

\vspace{1mm}
\noindent \textbf{Songyang Zhang} received the Ph.D. degree 
% in Department of Electrical and Computer Engineering 
from the University of California at Davis, Davis, CA, USA.
% , where he was a Postdoctoral Research Associate from 2021 to 2023. 
He is currently an Assistant Professor with the Department of Electrical and Computer Engineering at the University of Louisiana at Lafayette, Lafayette, LA, USA. 
% His current research interests include machine learning, signal processing, IoT intelligence and wireless communications. 
%He received the Best Paper Finalist in the 2020 IEEE International Conference on Image Processing, and was recognized as the Best TCSVT Reviewer of 2022 by IEEE Circuits and System Society.

\vspace{1mm}
\noindent\textbf{Guanying Chen} received the Ph.D. degree from The University of Hong Kong. He is currently an Associate Professor in the School of Cyber Science and Technology at Sun Yat-sen University (SYSU).
% His research interests are in computer vision and generative models.

\vspace{1mm}
\noindent \textbf{Wei Guo} received the Ph.D. degree from The Chinese University of Hong Kong, Shenzhen. He is currently a Postdoc researcher at The Hong Kong University of Science and Technology.

\vspace{1mm}
\noindent \textbf{Shenglun Lan} is with the School of Optoelectronic Engineering, Changchun University of Science and Technology, Changchun, China.

\vspace{1mm}
\noindent \textbf{Lexi Xu} received the Ph.D. degree from Queen Mary University of London, United Kingdom. He is now a professor level senior engineer at Research Institute, China Unicom. 
% He also serves as a professor (part-time) at BUPT. 
% He is also a China Unicom delegate in ITU, ETSI, 3GPP, CCSA. He also serves as a professor (part-time) at Beijing University of Posts and Telecommunications. He also serves as an industrial tutor (part-time) at Beijing Institute of Technology, Wuhan University, etc. He has applied for more than 50 patents, published 3 books, and edited 7 international conferences proceedings. His research interests include big data, self-organizing networks, satellite system, radio resource management in wireless system, etc.

\vspace{1mm}
\noindent \textbf{Xinzhou Cheng} is the professor level senior engineer, the head of network intelligent operation R\&D center at Research Institute, China Unicom. 
% He is also a professor (part-time) at Beijing University of Posts and Telecommunications (BUPT). 
% He received M.S. degree from Beijing University of Posts and Telecommunications in 2004. His research interests include telecom big data, network planning \& optimization, network intelligent operation. 

\vspace{1mm}
\noindent \textbf{Xiongyan Tang} received the Ph.D. degree from Beijing University of Posts and Telecommunications, Beijing, China. He is the chief scientist of China Unicom Research Institute, the vice Dean of China Unicom Research Institute. 
% He is also a professor in BUPT.
% , the state candidate of Millions of Talents Project in the New Century, a member of the Telecom Technology Committee of Ministry of Industry and Information Technology.  From 1994 to 1997, he conducted research of high-speed optical communications in Singapore and Germany. 
% He received his PhD degree in telecom engineering from Beijing University of Posts and Telecommunications in 1994. 
% Since 1998, he has been working on technology management in telecom operators in China.  He has published more than 150 technical papers. His professional fields include broadband communications, optical transmission, IP networks, SDN/NFV, telecom big data, new generation networks and Internet of Things.

% \vspace{2mm}
% \noindent \textbf{Shuguang Cui} received the Ph.D. in electrical engineering from Stanford University, Stanford, CA, USA, in 2005. Afterwards, he has been working as Assistant, Associate, Full, Chair Professor in Electrical and Computer Engineering with the University of Arizona, Tucson, AZ, USA; Texas A\&M University, College Station, TX, USA; University of California, Davis, CA, USA; and The Chinese University of Hong Kong (CUHK), Shenzhen, China, respectively.

\vspace{1mm}
\noindent \textbf{Shuguang Cui} received the Ph.D. degree from Stanford University, Stanford, CA, USA. He is currently the X.Q. Deng Presidential Chair Professor in the School of Science and Engineering at The Chinese University of Hong Kong, Shenzhen, China.

\end{document}